\documentclass{article}

\usepackage{arxiv}

\usepackage[utf8]{inputenc} 
\usepackage[T1]{fontenc}    
\usepackage{url}            
\usepackage{booktabs}       
\usepackage{amsfonts}       
\usepackage{nicefrac}       
\usepackage{microtype}      
\usepackage{graphicx}
\usepackage[square,numbers]{natbib}

\usepackage{amssymb}
\usepackage{amsmath}
\usepackage{tabularx}
\usepackage{mathtools}
\usepackage{bbold}
\usepackage{longtable}
\usepackage{makecell}
\usepackage{rotating}
\usepackage{hyperref}

\title{Enhanced Universal Kriging \linebreak for Transformed Input Parameter Spaces}

\date{} 					

\author{\href{https://orcid.org/0000-0002-3338-8406}{\includegraphics[scale=0.06]{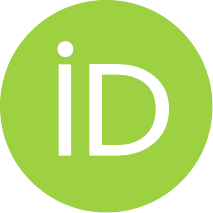}\hspace{1mm}Matthias Fischer}\\
	Institute of Engineering Mechanics\\
	Karlsruhe Institute of Technology\\
	76131 Karlsruhe, Germany\\
	\texttt{matthias.fischer@kit.edu} \\
	\And
	\href{https://orcid.org/0000-0003-4599-4531}{\includegraphics[scale=0.06]{orcid.pdf}\hspace{1mm}Carsten Proppe} \\
	Institute of Engineering Mechanics\\
	Karlsruhe Institute of Technology\\
	76131 Karlsruhe, Germany
}

\renewcommand{\headeright}{ }
\renewcommand{\undertitle}{ }
\renewcommand{\shorttitle}{Enhanced UK for Transformed Input Parameter Spaces}

\hypersetup{
pdftitle={Enhanced UK for Transformed Input Parameter Spaces},
pdfauthor={Matthias Fischer, Carsten Proppe},
}

\begin{document}
\maketitle

\begin{abstract}
	With computational models becoming more expensive and complex, surrogate models have gained increasing attention in many scientific disciplines and are often necessary to conduct sensitivity studies, parameter optimization etc. In the scientific discipline of uncertainty quantification (UQ), model input quantities are often described by probability distributions. For the construction of surrogate models, space-filling designs are generated in the input space to define training points, and evaluations of the computational model at these points are then conducted. The physical parameter space is often transformed into an i.\,i.\,d. uniform input space in order to apply space-filling training procedures in a sensible way. Due to this transformation surrogate modeling techniques tend to suffer with regard to their prediction accuracy. Therefore, a new method is proposed in this paper where input parameter transformations are applied to basis functions for universal kriging. To speed up hyperparameter optimization for universal kriging, suitable expressions for efficient gradient-based optimization are developed. Several benchmark functions are investigated and the proposed method is compared with conventional methods.
\end{abstract}

\textit{Manuscript accepted in: Probabilistic Engineering Mechanics on June 27, 2023\\
https://doi.org/10.1016/j.probengmech.2023.103486}\\
\\
\copyright 2023. This manuscript version is made available under the CC-BY-NC-ND 4.0 license\\ https://creativecommons.org/licenses/by-nc-nd/4.0/

\section{Introduction}

	Surrogate modeling methods are widely used in various scientific disciplines, such as engineering, meteorology or physics. Simulation runs for complex computational models are often computationally expensive. Surrogate modeling techniques provide the opportunity to construct a surrogate model with relatively few evaluations of the computational model.\\
	\\
	For instance, in meteorology, the German Meteorological Service (DWD) currently uses a global grid with a mesh size of $13$\,km to achieve sufficiently accurate weather predictions. Even at these high resolutions, many atmospheric processes occur on a sub-grid scale and still need to be parameterized, since they cannot be explicitly resolved. Consequently, these models are computationally expensive to run which makes further studies, such as parameter identification or sensitivity studies, infeasible.\\
	\\
	By using surrogate models, a relationship between model inputs and outputs can be determined with relatively low computational cost. Model parameters of the computational model are often considered as input parameters of the surrogate model in the context of uncertainty quantification (UQ). Epistemic uncertainties can arise from a lack of knowledge of these parameters. These uncertainties are often described by probability distributions that are determined based on measurements, expert knowledge or parameter identification studies. Before applying surrogate modeling techniques, input parameters are often transformed to independent and identically distributed (i.\,i.\,d.) uniform variables. In this unit hypercube, space-filling designs, such as Latin hypercube sampling, can be applied in a meaningful way. However, prediction accuracy of surrogate models may suffer from such transformations as will be shown in this paper. The aim of this work is to develop a strategy to overcome such prediction accuracy losses as a result of input space transformations.\\
	\\
	Although the exact relationship between input and output quantities is not known in advance, it may be possible to assume certain underlying trends in order to improve prediction accuracy. Such trends can be incorporated as basis functions by various surrogate modeling techniques, e.\,g. universal kriging. Usually, simple basis functions (often low order polynomials) are used for the sake of simplicity and to encounter the risk of overfitting. In a high dimensional input space, higher order multivariate polynomials up to a certain polynomial degree would lead to an excessive increase in the number of polynomials, commonly known as the "curse of dimensionality". To overcome this problem, methods for polynomial basis selection can be applied, such as least-angle regression. \citet{PCEkriging} proposed the method \textit{LARS-Kriging-PC modeling} where explicit multivariate basis functions for universal kriging are obtained by least-angle regression based on polynomial chaos expansions. The purpose of the method is to select polynomials that bring the most relevant information to the kriging model. The results show that the benefits of both Gaussian process regression and polynomial chaos expansion are combined. The concept is similar to \textit{blind kriging} where the underlying trend is identified from data using a Bayesian variable selection technique (\citet{blindkriging}). As with the case of \textit{LARS-Kriging-PC modeling}, a set of candidate basis functions has to be defined beforehand. In both methods, the choice of polynomials is not based on prior knowledge of the problem, but on the regression technique itself. However, \citet{kriging_oakley} emphasizes that the choice of basis functions should be chosen to incorporate any beliefs regarding the problem, e.\,g. the physical evolution of the output variable with respect to the input parameters (see e.\,g. \citet{PCEkriging}).\\
	\\
	None of the above mentioned methods consider the input parameter transformations to the i.\,i.\,d. uniform input space which is performed in this study. However, the transformation may highly impact the shape of the model function. Although it is straightforward to define simple polynomial basis functions in the i.\,i.\,d. uniform input space, i.\,e. the space in which the surrogate model is built, this is not the most meaningful choice. More precisely, defining basis functions in the original, physical parameter space is considered more sensible. In this paper, a method for defining basis functions for universal kriging is proposed that takes input parameter transformations into account. In particular, isoprobabilistic transformation is applied to the construction of basis functions.\\
	\\
	In this study it will be assumed that stationarity holds in the i.\,i.\,d. uniform space in reference to the uniform density of training points in this space. This can generally be considered to be a reasonable first approach if nothing is known about the physical problem. Accordingly, this corresponds to non-stationary basis functions in the physical input space. The idea of finding transformations between spaces so that stationarity and isotropy holds in the deformed space was already used by \citet{sampson} where thin-plate splines were applied to achieve such a mapping. \citet{ohagan_spacial} enhanced this method by using a Bayesian approach where the mapping is described by a function with a Gaussian process prior. Both methods have been developed for a setting with only two input dimensions in the context of geostatistics.\\
	\\
	A related approach of incorporating non-stationary kernel functions has been proposed by \citet{Gibbs} and a simplified version has been demonstrated by \citet{Xiong}, where a density function is constructed that aims at describing the smoothness of function value changes with respect to input parameters. The density function is then used to describe a non-linear mapping to an input space where uniform smoothness and therefore a stationary kernel function can be assumed. Using this mapping, non-stationary kernel functions can be defined in the original input space. The difference compared to this paper is that in the approach by \citet{Gibbs} the transformation is based on the estimated smoothness of function values, whereas here it is based on the joint PDF of input parameters and thus the density of training points.\\
	\\
	The core of Gaussian process regression methods lies in the estimation of hyperparameters. These are generally determined by likelihood maximization. With a high number of training points and input dimensions, the estimation of hyperparameters may be computationally expensive. Therefore, gradient-based hyperparameter estimation for Gaussian process regression is frequently applied to speed up hyperparameter optimization. The derivative of the log marginal likelihood with respect to the hyperparameters is determined for universal kriging in this work.\\
	\\
	The remainder of this paper is organized as follows. In \autoref{sec:quantile} applied methods are depicted. First, the applied training procedure and fundamentals of Gaussian process regression are presented. Then, a technique for the construction of basis functions for transformed input parameter spaces is proposed. Suitable expressions for gradient-based hyperparameter optimization for universal kriging are developed. Model validation measures are specified. In \autoref{sec:test} the model setup and the set of benchmark problems for investigation of proposed methods are presented. Validation results are shown in \autoref{sec:results} an discussed in \autoref{sec:discussion}. Finally, \autoref{sec:conclude} concludes this work.

\section{Universal kriging with basis functions for transformed input parameter spaces}\label{sec:quantile}
	
	In this section, the applied training procedure, simple kriging and universal kriging are briefly presented. A new way of defining basis functions for transformed input parameter spaces is then proposed. Equations for gradient-based hyperparameter optimization for universal kriging are developed. Finally, the applied model validation strategy is described.\\

	\subsection{Training procedure}\label{sec:training}
	
		In this study, problems with non-uniform input parameter distributions are considered. In terms of training strategies, it is reasonable to generate a space-filling design that takes the PDFs of input parameters into account, i.e. to generate more training points in regions associated with higher probability and vice versa. This generally yields higher overall prediction accuracy, because the resulting surrogate model is then more accurate in regions associated with higher probabilities. In particular, lower validation errors are obtained, if the model is validated by Monte Carlo sampling with respect to parameter PDFs. \citet{non-uniform-design} described a procedure to find such space-filling designs. However, surrogate modeling methods, such as Gaussian process regression, may struggle with inhomogeneous density of training points in the input space, because optimal covariance parameters are usually strongly dependent on the density of training points. However, if specific problems are considered where a higher accuracy is desired in certain regions of the input space, as it is the case in reliability analysis (see e.\,g. \citet{releng}), adjusted methods are preferred that assign a higher training point density to such regions.\\
		\\
		Therefore, the focus of this study is the case where input parameter spaces are transformed to i.\,i.\,d. uniform parameter spaces. Isoprobabilistic transformation, in particular Rosenblatt transformation \cite{rosenblatt}, is applied to transform the physical input vector $\mathbf x = (x_1, x_2 \ldots x_p)^\top$ to the i.\,i.\,d. uniform input vector $\mathbf u = (u_1, u_2 \ldots u_p)^\top$ with input space dimension $p$.	This makes the application of space-filling designs more sensible, because a homogeneous space-filling design can then be obtained in the uniform input space. Thus, the training point density induces a model accuracy which corresponds to the probability of parameter values. Furthermore, in the case of highly different scales between the input parameters, surrogate modeling methods may suffer without such a transformation.  Let $\hat {\mathbf x} = (\hat x_1, \hat x_2 \ldots \hat x_p)^\top$ be a random vector in the physical input space with joint cumulative distribution function $F_{\hat {\mathbf x}} (\mathbf x)$. The Rosenblatt transformation ${\mathbf u} = \mathcal T_\text{ros} ( {\mathbf x})$ is then given by
		\begin{align*} 
			u_{1} &=P\left\{\hat x_{1} \leq x_{1}\right\}=F_{\hat x_1}\left(x_{1}\right) \\ 
			u_{2} &=P\left\{\hat x_{2} \leq x_{2} \mid \hat x_{1}=x_{1}\right\}=F_{\hat x_2|\hat x_1}\left(x_{2} \mid x_{1}\right) \\ & \vdots \\ 
			u_{p} &=P\left\{\hat x_{p} \leq x_{p} \mid \hat x_{p-1}=x_{p-1} \ldots \hat x_{1}=x_{1}\right\}=F_{\hat x_p|\hat x_{p-1}\ldots \hat x_{1}}\left(x_{p} \mid x_{p-1} \ldots x_{1}\right),
		\end{align*}
		where $F_{\,\Box}()$ denote respective conditional cumulative distribution functions. It can be shown that transformed random vector $\hat {\mathbf u} = (\hat u_1, \hat u_2 \ldots \hat u_p)^\top = \mathcal T_\text{ros} (\hat {\mathbf x}) $ is then i.\,i.\,d. uniformly distributed on the $p$-dimensional unit hypercube. $F_{\,\Box}()$ can be determined from the joint PDF $f_{\hat{\mathbf x}}(\mathbf x)$ which is assumed to be known in this study (see e.\,g. \citet{rosenblatt_pdf}).\\
		\\
		Maximin Latin hypercube sampling (\citet{maximin_lhd}) for generating a set of $n$ training points in the i.\,i.\,d. uniform parameter space $\mathbf U = \{\mathbf u_i, \, i = 1\ldots n\}$ is applied, where $\mathbf U$ corresponds to the set of training points in the physical parameter space $\mathbf X = \{\mathbf x_i, \, i = 1\ldots n\}$ such that $\{\mathbf u_i = \mathcal T_\text{ros}(\mathbf x_i),\, i = 1\ldots n\}$. Evaluations of the computational model $f(\mathbf x_i)$ are conducted for all training points resulting in the output vector $\mathbf y = \{y_i=f(\mathbf x_i),  \,i=1\ldots n\}$.\\
		\\
		The aim of kriging is to build a surrogate model $\mathcal M$ for a scalar model output $y$, i.\,e. a quantity of interest (QoI), based on the experimental design and model evaluations $\{\mathbf U, \, \mathbf y\}$. Note that all equations in the context of kriging are represented with respect to i.\,i.\,d. uniform variables $\mathbf u$ instead of physical variables $\mathbf x$, because surrogate models are built in the i.\,i.\,d. uniform parameter space in this study. Prediction mean and prediction variance at a set of input points $\mathbf U_\star = \{\mathbf {u_\star}_{\,i}, \, i = 1\ldots l\}$ are to be determined. \\
		
	\subsection{Simple kriging}\label{sec:ordkrig}
		For simple kriging, a zero-mean Gaussian process
		\begin{equation}\label{eq:sk}
			g_{\text{SK}}(\mathbf u) \sim \mathcal{GP}(0,k(\mathbf u,\mathbf u'))
		\end{equation}
		with covariance function $k(\mathbf u,\mathbf u')$, also known as kernel function, is assumed as prior. The kernel function describes the dependence structure between values of the stochastic process at different points, usually depending on their distance.\\
		\\
		Here, the anisotropic form of the radial-basis function
		\begin{equation}
			k(\mathbf u,\mathbf u') = \theta_0 \, \, \text{exp}\left(-\sum_{i=1}^p\left(\frac{|u_i - u_i'|}{\theta_i}\right)^2\right) \label{eq:kernel}
		\end{equation}
		with respect to hyperparameters $\boldsymbol \theta = \{\theta_i, i = 0\ldots p\}$ is used to allow for different smoothness between input dimensions. Furthermore, i.\,i.\,d. Gaussian noise with variance $\sigma_n^2$ is added to allow for aleatoric uncertainties in the simulations of the computational model.\\
		\\
		Let
		\begin{alignat*}{5}
			&\mathbf K = &&\{K_{ij} &&= \, k(\mathbf u_i, \mathbf u_j),\, &&i = 1\ldots n,\quad &&j = 1\ldots n\} \, ,\\
			&\mathbf k = &&\{k_{ij} &&= \,k(\mathbf u_i\, ,\,  {\mathbf u_\star \,}_j ),\, &&i = 1\ldots n,\, &&j = 1\ldots l\}\quad \text{and}\\
			&\boldsymbol \sigma_0^2 = &&\{{\sigma_0^2}_j &&=\, k( {\mathbf u_\star\,}_j,   {\mathbf u_\star\,}_j),\quad &&j = 1\ldots l\}&&
		\end{alignat*} 
		be the vectors and matrices of kernel function evaluations at training points $\mathbf U$ and prediction points $\mathbf U_\star$, respectively.\\
		\\
		The best linear unbiased predictor (BLUP) and its prediction variance for the set of prediction points $\mathbf U_\star$ under the assumptions of simple kriging, as shown by \citet{rasmussen}, are then
		\begin{align*}
			\boldsymbol{\mathcal{M}}(\mathbf U_\star) &=  \mathbf k^\top \mathbf K_y ^{-1} \mathbf y \\
			\boldsymbol \sigma^2(\mathbf U_\star) & = \boldsymbol \sigma_0^2 - \mathbf k^\top \mathbf K_y^{-1} \mathbf k
		\end{align*}
		with $\mathbf K_y = \mathbf K + \sigma_n^2 \mathbb{1}$. The hyperparameters $\boldsymbol \theta$ and noise parameter $\sigma_n$ are determined by maximum likelihood estimation.\\
	
	\subsection{Universal kriging}\label{sec:unikrig}
		
		The theory of universal kriging was introduced by \citet{universal_kriging} in the field of geostatistics. A prior
		\begin{equation}\label{eq:uk}
			g_{\text{UK}}(\mathbf u)=g_{\text{SK}}(\mathbf u)+\mathbf h(\mathbf u)^\top \boldsymbol \beta
		\end{equation}
		is used, with zero-mean Gaussian process $g_{\text{SK}}(\mathbf u)$ (\autoref{eq:sk}), vectors of known basis functions $\mathbf h(\mathbf u) = \{h_j(\mathbf u),\, j = 1\ldots q\}$ and  coefficients $\boldsymbol \beta = \{\beta_j, \, j = 1\ldots q\}$. Coefficients $\boldsymbol \beta$ are unknown, but not required to be specified for the computation (see \citet{rasmussen}).\\
		\\
		Let 
		\begin{alignat}{5}
			&\mathbf H = &&\{H_{ij} &&= h_i(\mathbf u_j),\quad&& i = 1\ldots q,\quad &&j = 1\ldots n\}\quad \text{and}\label{eq:H1}\\
			&\mathbf H_\star = &&\{{H_\star}_{\,ij} &&= h_i(\mathbf {u_\star}_{\,j}),\quad &&i = 1\ldots q,\, &&j = 1\ldots l\}\label{eq:H2}
		\end{alignat} 
		be the matrices of basis function evaluations at training points $\mathbf U$ and prediction points $\mathbf U_\star$, respectively.\\
		\\
		The best linear unbiased predictor (BLUP) and its prediction variance for the set of prediction points $\mathbf U_\star$ under the assumptions of universal kriging, as shown by \citet{rasmussen}, are then
		\begin{align}
			\boldsymbol{\mathcal{M}}(\mathbf U_\star) &= \mathbf H_\star^\top \boldsymbol \mu + \mathbf k ^\top \mathbf K_y^{-1} (\mathbf y - \mathbf H^\top \boldsymbol \mu)\label{eq:uk_sol1}\\
			\boldsymbol \sigma^2(\mathbf U_\star) & = \boldsymbol \sigma_0^2 - \mathbf k^\top \mathbf K_y^{-1} \mathbf k + \mathbf R^\top (\mathbf H\mathbf K_y^{-1}\mathbf H^\top)^{-1}\mathbf R\label{eq:uk_sol2}
		\end{align}
		with $\boldsymbol \mu = (\mathbf H \mathbf K_y ^{-1}\mathbf H^\top)^{-1}\mathbf H \mathbf K_y ^{-1} \mathbf y$\, and \, $\mathbf R=\mathbf H_\star - \mathbf H \mathbf K_y^{-1} \mathbf k \, $.\\
		\\
		In the case of a constant scalar basis function $h(\mathbf u) = 1$, \autoref{eq:uk_sol1} and \autoref{eq:uk_sol2} become
		\begin{align*}
			\boldsymbol{\mathcal{M}}(\mathbf U_\star) &= \mu \, \mathbf I + \mathbf k ^\top \mathbf K_y^{-1} (\mathbf y - \mu \, \mathbf I)\\
			\boldsymbol \sigma^2(\mathbf U_\star) & = \boldsymbol \sigma_0^2 - \mathbf k^\top \mathbf K_y^{-1} \mathbf k + \mathbf R^\top (\mathbf H\mathbf K_y^{-1}\mathbf H^\top)^{-1}\mathbf R 
		\end{align*}
		where $\mu = \frac{\mathbf I^\top \mathbf K_y ^{-1} \mathbf y}{\mathbf I^\top \mathbf K_y ^{-1} \mathbf I}$. This case is known as ordinary kriging, i.\,e. kriging with unknown constant mean.

	\subsection{Basis functions for transformed input parameter spaces}\label{sec:basis_functions}
		
\begin{figure*}[p!]
			\includegraphics[width=14cm]{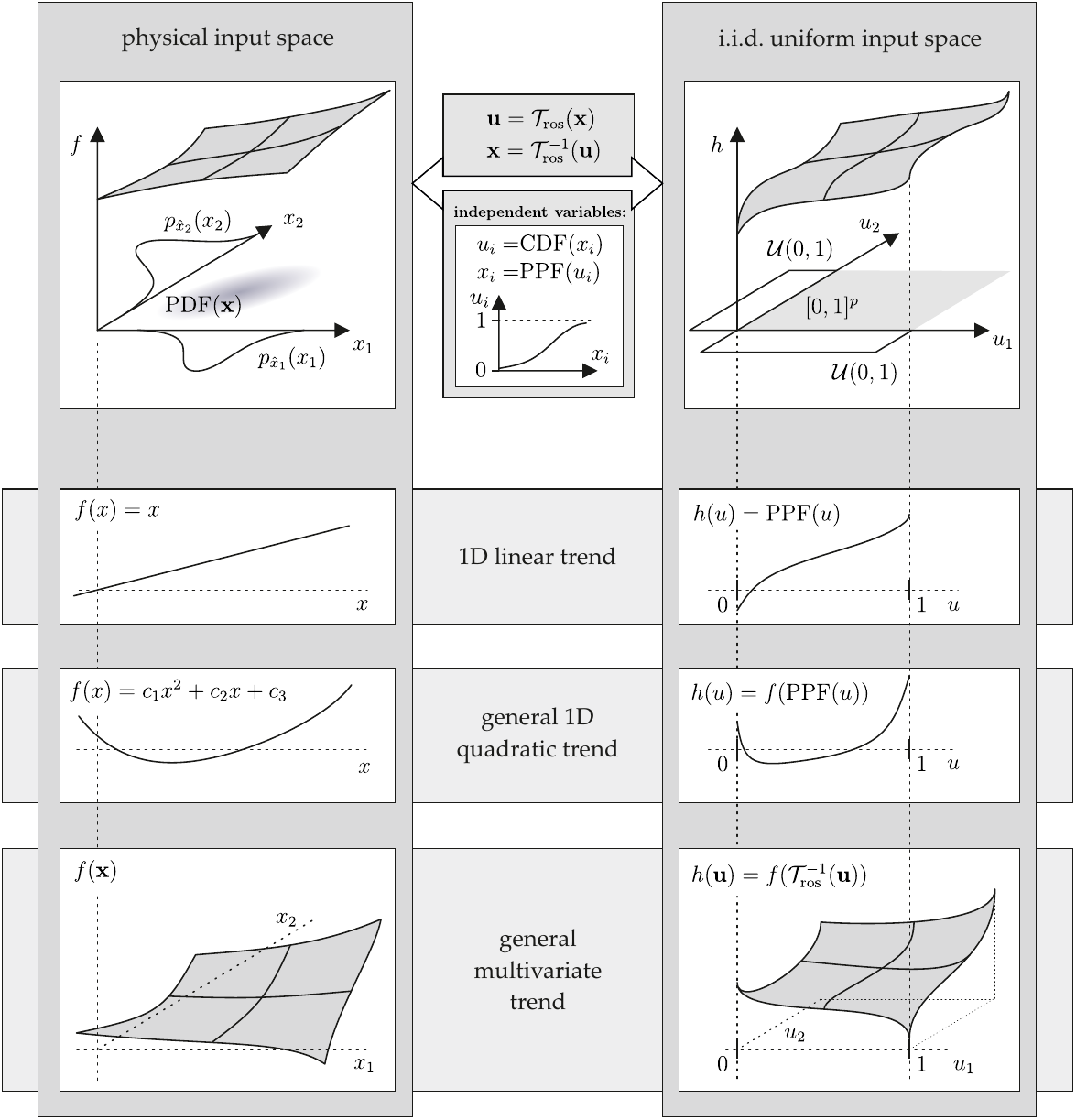}
			\caption{Visualization of model functions (top) and selected trend functions (three rows below) in the physical (left) and i.\,i.\,d. uniform (right) parameter space. The Rosenblatt transformation $\mathcal T_{\text{ros}}$ (quantile functions PPF in case of independent variables) defines the relationship between both parameter spaces.}
			\label{fig:illustration}
		\end{figure*}

		In many cases, a zero or constant mean function for the Gaussian process is sufficient which corresponds to scalar basis functions $\mathbf h(u) = 0$ or $\mathbf h(u) = 1$ in \autoref{eq:uk}, respectively. However, incorporation of more sophisticated basis functions may significantly improve prediction accuracy. \citet{PCEkriging} proposed the method \textit{LARS-Kriging-PC modeling} where explicit basis functions for universal kriging are obtained by least-angle regression based on polynomial chaos expansions. A sparse polynomial basis is generated to tackle the curse of dimensionality. Similarly, for \textit{blind kriging}, the underlying trend is identified from data using a Bayesian variable selection technique (\citet{blindkriging}). In both cases, basis functions are selected from a set of candidate functions based on the surrogate modeling technique itself rather than physical knowledge about the problem.\\
		\\
		In particular, the selection of basis functions becomes more important when using transformed input parameter spaces. In this study, transformation of the physical input space to an i.\,i.\,d. uniform input space is carried out as described in \autoref{sec:training}. Kriging techniques are then applied in the i.\,i.\,d. uniform input space. Although it is then straightforward to define simple polynomial basis functions in the i.\,i.\,d. uniform input space, this is not the most meaningful choice. Defining basis functions in the original, physical parameter space is considered to be more sensible. This is because physical relationships that are attempted to be captured by trend functions are assumed to be related more directly to physical input parameters than to transformed i.\,i.\,d. uniform input parameters. Therefore, it is proposed to define polynomial basis functions $f(\mathbf x)$ in the original, physical input space $\mathbf x$ and express them as functions $h(\mathbf u) = f(\mathcal T_{\text{ros}}^{-1} ({\mathbf u}))$ in the i.\,i.\,d. uniform input space ${\mathbf u}$ by means of the inverse Rosenblatt transformation $\mathcal T_{\text{ros}}^{-1} $ (see \autoref{sec:training}). In the one-dimensional case, the transformation becomes the quantile function or percent-point-function $x = \text{PPF}(u)$ (inverse cumulative distribution function) of the input parameter. The transformed basis function is then $h(u) = f(\text{PPF}(u))$. In case of the particular linear basis function $f(x)=x$, the transformed basis function results in the quantile function $h(u) = \text{PPF}(u)$. If there is no correlation between the input parameters, i.e. $\{\rho_{\hat x_i, \hat x_j}=0 \quad \forall i,j \in \{1,\ldots ,p\}, \, i\neq j\}$, the transformation reduces to independent functions $\{h(u_i)=f(\text{PPF}(u_i)),\, i = 1\ldots p\}$ in all input parameters.\\
		\\
		\autoref{fig:illustration} illustrates the transformation of basis functions. Transformed basis functions $\mathbf h({\mathbf u})=\{h_j( {\mathbf u}), \, j=1\ldots q\}$ can then be used as basis functions in the universal kriging method (\autoref{eq:uk}), i.\,e. for computing $\mathbf H$ and $\mathbf H_\star$ (\autoref{eq:H1} and \autoref{eq:H2}).\\
		\\
	
	\subsection{Link to non-stationary kernel functions}\label{sec:nonstat}
	
		In this paper, the procedure of building surrogate models is carried out in the i.\,i.\,d. uniform parameter space. However, by using the Rosenblatt transformation, uniform input variables can be transformed into the physical parameter space as described in \autoref{sec:training}. All equations from \autoref{sec:ordkrig} and \autoref{sec:unikrig} can thus be expressed with respect to physical parameters $\mathbf x$. In the case of universal kriging, the original non-transformed definition of trend functions in the physical parameter space are used instead of the transformed basis functions in the i.\,i.\,d. uniform parameter space according to \autoref{sec:basis_functions}.\\
		\\
		In particular, the description of equations results in non-stationary kernel functions, because the kernel functions are defined as stationary with respect to i.\,i.\,d. uniform variables. The anisotropic formulation of the radial-basis function kernel (\autoref{eq:kernel}) thus results in
		\begin{equation*}
			k(\mathbf x,\mathbf x') =  \theta_0 \, \, \text{exp}\left(-\sum\limits_{i=1}^p\left(\frac{|\mathcal T_\text{ros}(\mathbf x)_i - \mathcal T_\text{ros}(\mathbf x')_i|}{\theta_i}\right)^2\right). 
		\end{equation*}
		It is emphasized that the definition of stationary kernel functions in the i.\,i.\,d. uniform parameter space is meaningful instead of using stationary kernel functions in the physical parameter space to account for the non-homogeneous density of training points. According to \autoref{sec:training}, the density of training points in the physical parameter space corresponds to the joint PDF of input parameters. Training points are sparse in the tails of the PDF and a larger length scale of the kernel function in these regions is desirable, whereas in regions with a higher density of training points, i.\,e. closer to the mean value, a smaller length scale is sensible.\\

	\subsection{Gradient-based hyperparameter estimation for universal kriging}\label{sec:hyperpar}
		
		Hyperparameter optimization for universal kriging can be computationally intensive, particularly with many input dimensions and a large number of training points. Optimization of the hyperparameters is most commonly conducted by maximizing the marginal likelihood which is a measure for the model evidence. However, other methods exist, such as maximization of the \textit{pseudo-likelihood} which is obtained by cross-validation (see \citet{rasmussen} for further discussion), which are not considered in this work. Preferably, gradient-based hyperparameter optimization with incorporation of gradient information is conducted to reduce computation time. Therefore, the gradient of the log marginal likelihood with respect to the hyperparameters $\boldsymbol \theta$ has to be determined. \\
		\\
		The log marginal likelihood for simple kriging is given as
		\begin{equation}
			\log p(\mathbf y | \mathbf U, \boldsymbol{\theta}) = -\frac{1}{2}\mathbf y^\top \mathbf K_y^{-1} \mathbf y - \frac{1}{2} \log | \mathbf K_y | - \frac{p}{2} \log 2\pi. \label{eq:log_lik_0}
		\end{equation}
		\\
		Since only $\mathbf K_y$ depends on $\boldsymbol \theta$, an expression for the gradient of the log marginal likelihood is aimed that only depends on $\frac{\partial \mathbf K_y}{\partial \boldsymbol \theta}$. If the gradient of $\mathbf K_y$ is available, as it is the case for widely used kernel functions, the gradient of the log marginal likelihood can be determined.\\
		\\
		For simple kriging, the gradient of the log marginal likelihood by the use of matrix identities in \autoref{sec:appendix} becomes
		\begin{align*} \frac{\partial}{\partial \theta_{l}} \log p(\mathbf{y} | \mathbf U, \boldsymbol{\theta}) &=\frac{1}{2} \mathbf{y}^{\top} \mathbf K_y^{-1} \frac{\partial \mathbf K_y}{\partial \theta_{l}} \mathbf K_y^{-1} \mathbf{y}-\frac{1}{2} \operatorname{tr}\left(\mathbf K_y^{-1} \frac{\partial \mathbf K_y}{\partial \theta_{l}}\right) \\ &=\frac{1}{2} \operatorname{tr}\left(\left(\boldsymbol{\alpha} \boldsymbol{\alpha}^{\top}-\mathbf K_y^{-1}\right) \frac{\partial \mathbf K_y}{\partial \theta_{l}}\right) 
		\end{align*}
		with $\boldsymbol{\alpha}=\mathbf K_y^{-1} \mathbf{y} = \mathbf L_K^{-\top} (\mathbf L_K^{-1}\mathbf y)$ and $\mathbf L_K = \text{cholesky} (\mathbf K_y)$.\\
		\\
		Using the Einstein summation convention, the derivative of the log marginal likelihood given $\boldsymbol \alpha$ can be expressed as
		\begin{align}
			\frac{\partial}{\partial \theta_l}\log p(\mathbf y | \mathbf U, \boldsymbol{\theta}) &= \frac{1}{2} \left(\left(\alpha_{i} \, \alpha_{j}  -\delta_{im} {\left[\mathbf K_y^{-1}\right]}_{mj}\right) \frac{\partial {K_y}_{\,ji}}{\partial \theta_l}\right)\label{eq:log0}
		\end{align}
		with Kronecker delta $\delta_{im}$.\\
		\\
		In the following, the derivation of the gradient of the log marginal likelihood for universal kriging is carried out, where suitable abbreviations for efficient computation are introduced.\\
		\\
		The log marginal likelihood for universal kriging, as shown by \citet{rasmussen}, is
		\begin{equation}
			\log p(\mathbf y | \mathbf U, \boldsymbol{\theta}) = - \frac{1}{2} \mathbf y^\top \mathbf K_y^{-1} \mathbf y + \frac{1}{2} \mathbf y^\top \mathbf C \mathbf y - \frac{1}{2} \log |\mathbf K_y| - \frac{1}{2} \log |\mathbf A| - \frac{p-m}{2} \log 2\pi, \label{eq:log_lik_1}
		\end{equation}
		where $\mathbf A=\mathbf H \mathbf K_y^{-1} \mathbf H^\top$, $\mathbf C=\mathbf K_y^{-1} \mathbf H^\top \mathbf A^{-1}\mathbf H \mathbf K_y^{-1}$ and $m=\text{rank}(\mathbf H^\top)$.\\
		\\
		Inserting $\mathbf A$ and $\mathbf C$ in \autoref{eq:log_lik_1} results in
		\begin{align*}
			\log p(\mathbf y | \mathbf U, \boldsymbol{\theta}) = &- \frac{1}{2} \mathbf y^\top \mathbf K_y^{-1} \mathbf y + \frac{1}{2} \mathbf y^\top \left(\mathbf K_y^{-1} \mathbf H^\top (\mathbf H \mathbf K_y^{-1} \mathbf H^\top)^{-1}\mathbf H \mathbf K_y^{-1}\right) \mathbf y \\
			&- \frac{1}{2} \log |\mathbf K_y| - \frac{1}{2} \log |\mathbf H \mathbf K_y^{-1} \mathbf H^\top| - \frac{n-m}{2} \log 2\pi\\
			=& - \frac{1}{2} \mathbf y^\top \boldsymbol \alpha + \frac{1}{2} \mathbf y^\top \boldsymbol \gamma \boldsymbol \eta \boldsymbol \alpha \\
			&- \frac{1}{2} \log |\mathbf K_y| - \frac{1}{2} \log |\mathbf H \boldsymbol \gamma| - \frac{n-m}{2} \log 2\pi,
		\end{align*}
		where $\boldsymbol \alpha$,  $\boldsymbol \gamma$ and  $\boldsymbol \eta$ are defined as
		\begin{align}
			\boldsymbol \alpha &= \mathbf K_y^{-1} \mathbf y = \mathbf L_K^{-\top} (\mathbf L_K^{-1}\mathbf y),\label{eq:alpha1}\\
			\boldsymbol \gamma &= \mathbf K_y^{-1} \mathbf H^\top = \mathbf L_K^{-\top} (\mathbf L_K^{-1}\mathbf H^\top) \quad \text{and} \label{eq:gamma1}\\
			\boldsymbol \eta &= (\mathbf H \mathbf K_y^{-1} \mathbf H^\top)^{-1}\mathbf H = \mathbf L_{\eta}^{-\top} (\mathbf L_{\eta}^{-1} \mathbf H) \label{eq:eta1}
		\end{align}
		with $\mathbf L_K = \text{cholesky} (\mathbf K_y)$ and $\mathbf L_{\eta} = \text{cholesky} (\mathbf H \mathbf K_y^{-1} \mathbf H^\top)$. Cholesky decomposition is applied in order to efficiently determine the inverse.\\
		\\
		Taking the derivative with respect to $\boldsymbol \theta$ by the use of matrix identities in \autoref{sec:appendix} results in
		\begin{align*}
			\frac{\partial}{\partial \theta_l}&\log p(\mathbf y | \mathbf U, \boldsymbol \theta)\\
			= & \quad \frac{1}{2} \mathbf y^\top \mathbf K_y^{-1} \frac{\partial \mathbf K_y}{\partial \theta_l} \mathbf K_y^{-1} \mathbf y\\
			& - \frac{1}{2} \mathbf y^\top \left(\mathbf K_y^{-1} \frac{\partial \mathbf K_y}{\partial \theta_l} \mathbf K_y^{-1} \mathbf H^\top (\mathbf H \mathbf K_y^{-1} \mathbf H^\top)^{-1}\mathbf H \mathbf K_y^{-1}\right) \mathbf y \\
			& + \frac{1}{2} \mathbf y^\top \left(\mathbf K_y^{-1} \mathbf H^\top (\mathbf H \mathbf K_y^{-1} \mathbf H^\top)^{-1}\mathbf H \mathbf K_y^{-1} \frac{\partial \mathbf K_y}{\partial \theta_l} \mathbf K_y^{-1} \mathbf H^\top (\mathbf H \mathbf K_y^{-1} \mathbf H^\top)^{-1} \mathbf H \mathbf K_y^{-1}\right) \mathbf y\\
			& - \frac{1}{2} \mathbf y^\top \left(\mathbf K_y^{-1} \mathbf H^\top (\mathbf H \mathbf K_y^{-1} \mathbf H^\top)^{-1}\mathbf H \mathbf K_y^{-1} \frac{\partial \mathbf K_y}{\partial \theta_l} \mathbf K_y^{-1}\right) \mathbf y \\
			& - \frac{1}{2} \text{tr}\left(\mathbf K_y^{-1} \frac{\partial \mathbf K_y}{\partial \theta_l}\right) - \frac{1}{2} \text{tr} \left(-(\mathbf H \mathbf K_y^{-1} \mathbf H^\top)^{-1}\mathbf H \mathbf K_y^{-1}  \frac{\partial \mathbf K_y}{\partial \theta_l} \mathbf K_y^{-1}\mathbf H^\top\right).
		\end{align*}
		\\
		In the next step, symmetry and positive definiteness of $\mathbf K_y$ (by definition) and therefore symmetry of $\mathbf K_y^{-1}$ are taken into account. Furthermore, the matrix identity in \autoref{eq:spur} is used to simplify the result. With $\boldsymbol \alpha, \boldsymbol \eta$ and $\boldsymbol \gamma$, it follows
		\begin{align*}
			\frac{\partial}{\partial \theta_l}&\log p(\mathbf y | \mathbf U, \boldsymbol \theta) \\
			= & \quad \frac{1}{2} \boldsymbol \alpha ^\top \frac{\partial \mathbf K_y}{\partial \theta_l} \boldsymbol \alpha - \frac{1}{2} \boldsymbol \alpha ^\top \frac{\partial \mathbf K_y}{\partial \theta_l} \boldsymbol \gamma \boldsymbol \eta \boldsymbol \alpha + \frac{1}{2}\boldsymbol \alpha^\top \boldsymbol \eta^\top \boldsymbol \gamma^\top \frac{\partial \mathbf K_y}{\partial \theta_l} \boldsymbol \gamma \boldsymbol \eta \boldsymbol \alpha\\
			& - \frac{1}{2}\boldsymbol \alpha^\top \boldsymbol \eta^\top \boldsymbol \gamma^\top \frac{\partial \mathbf K_y}{\partial \theta_l} \boldsymbol \alpha - \frac{1}{2} \text{tr}\left(\mathbf K_y^{-1} \frac{\partial \mathbf K_y}{\partial \theta_l}\right) + \frac{1}{2} \text{tr} \left(\boldsymbol \gamma \boldsymbol \eta \mathbf K_y^{-1}  \frac{\partial \mathbf K_y}{\partial \theta_l}\right)\\
			= & \quad \frac{1}{2} \text{tr} \bigg( \boldsymbol \alpha ^\top \frac{\partial \mathbf K_y}{\partial \theta_l} \boldsymbol \alpha - \boldsymbol \alpha ^\top \frac{\partial \mathbf K_y}{\partial \theta_l} \boldsymbol \gamma \boldsymbol \eta \boldsymbol \alpha\\
			&\hspace{5cm}+ \boldsymbol \alpha^\top \boldsymbol \eta^\top \boldsymbol \gamma^\top \frac{\partial \mathbf K_y}{\partial \theta_l} \boldsymbol \gamma \boldsymbol \eta \boldsymbol \alpha - \boldsymbol \alpha^\top \boldsymbol \eta^\top \boldsymbol \gamma^\top \frac{\partial \mathbf K_y}{\partial \theta_l} \boldsymbol \alpha  \bigg)\\
			&- \frac{1}{2} \text{tr}\left(\mathbf K_y^{-1} \frac{\partial \mathbf K_y}{\partial \theta_l}\right) + \frac{1}{2} \text{tr} \left(\boldsymbol \gamma \boldsymbol \eta \mathbf K_y^{-1}  \frac{\partial \mathbf K_y}{\partial \theta_l}\right)\\
			= & \quad \frac{1}{2} \text{tr} \bigg(\bigg( \boldsymbol \alpha \boldsymbol \alpha ^\top  - \boldsymbol \gamma \boldsymbol \eta \boldsymbol \alpha \boldsymbol \alpha ^\top + \boldsymbol \gamma \boldsymbol \eta \boldsymbol \alpha \boldsymbol \alpha^\top \boldsymbol \eta^\top \boldsymbol \gamma^\top  \\
			&\hspace{5cm} - \boldsymbol \alpha \boldsymbol \alpha^\top \boldsymbol \eta^\top \boldsymbol \gamma^\top  - \mathbf K_y^{-1} + \boldsymbol \gamma \boldsymbol \eta \mathbf K_y^{-1} \bigg) \frac{\partial \mathbf K_y}{\partial \theta_l}\bigg).
		\end{align*}
		The abbreviations $\quad \boldsymbol \rho =  \boldsymbol \alpha \boldsymbol \alpha ^\top, \quad\boldsymbol \varepsilon = \boldsymbol \gamma \boldsymbol \eta\quad$ and $\quad\boldsymbol \xi = \boldsymbol \varepsilon \boldsymbol \rho$\quad are introduced. It follows
		\begin{align*}
			\frac{\partial}{\partial \theta_l}\log p(\mathbf y | \mathbf U, \boldsymbol \theta)  = \frac{1}{2} \text{tr} \left(\left( \boldsymbol \rho - \boldsymbol \xi  - \boldsymbol \xi ^\top + \boldsymbol \xi \boldsymbol \varepsilon ^\top  + ( \boldsymbol \varepsilon- \mathbb{1}) \mathbf K_y^{-1}\right) \frac{\partial \mathbf K_y}{\partial \theta_l}\right).
		\end{align*}
		Using the Einstein summation convention, the following calculation steps are required to determine the derivative of the log marginal likelihood given $\boldsymbol \alpha$, $\boldsymbol \gamma$ and $\boldsymbol \eta$ from \autoref{eq:alpha1},  \autoref{eq:gamma1} and \autoref{eq:eta1}:
		\begin{align}
			\rho_{ij} & = \alpha_{i} \, \alpha_{j}, \nonumber\\
			\varepsilon_{ij} &= \gamma_{ik}\, \eta_{kj}, \nonumber\\
			\xi_{il}&=\varepsilon_{ij}\,\rho_{jl},\nonumber\\
			\frac{\partial}{\partial \theta_l}&\log p(\mathbf y | \mathbf U, \boldsymbol{\theta}) \notag\\
			&= \frac{1}{2} \left(\left(\rho_{ij} - \xi_{ij}  - \xi_{ji} + \xi_{im} \varepsilon_{jm}  + (\varepsilon_{im}-\delta_{im}) {\left[\mathbf K_y^{-1}\right]}_{mj}\right) \frac{\partial {K_y}_{\,ji}}{\partial \theta_l}\right)\label{eq:log1}
		\end{align}
		with Kronecker delta $\delta_{im}$. \\
		\\
		The complexity of the computation of the gradient of the log marginal likelihood (\autoref{eq:log0} and \autoref{eq:log1}) is dominated by the inverse of matrix $\mathbf K_y$ which is of the computational complexity $\mathcal{O}(n^3)$. Once the inverse is determined, it can be used for the computation of all hyperparameters $\theta_l$. In contrast, computation of the gradient of the log marginal likelihood based on \autoref{eq:log_lik_0} and \autoref{eq:log_lik_1}, i.\,e. without using the gradient of matrix $\mathbf K_y$, is of the computational complexity $\mathcal{O}(p\cdot n^3)$, because the inverse of $\mathbf K_y$ has to be determined towards all input dimensions $p$. Gradient-based hyperparameter optimization with incorporation of gradient information is therefore more beneficial.\\
		\\
		
	\subsection{Model validation}\label{sec:valid}
		When surrogate models are constructed, it is essential to assess their quality based on model validation measures. Therefore, the generalization error
		\begin{align*}
			\text{SE} &= \mathbb E \left[(f(\mathbf{ \hat x}) - \mathcal M(\hat {\mathbf x}))^2 \right]
		\end{align*}
		is considered, where $\mathbb E[\,\cdot\,]$ is the mathematical expectation operator. It describes the squared difference between original physical model $f$ and surrogate model prediction $\mathcal M$ and is therefore denoted as SE (\textit{squared error}). It can be expressed as
		\begin{align}
			\text{SE} &=\int_{\mathcal D_{\mathbf{ \hat x}}} (f(\mathbf x) - \mathcal M(\mathbf x))^2 \, f_{\mathbf{ \hat x}}(\mathbf x) \text{d} \mathbf x \nonumber \\
			&=\int_{\mathcal D_{\mathbf{ \hat u}}} (f(\mathbf u) - \mathcal M(\mathbf u))^2 \, f_{\mathbf{ \hat u}}(\mathbf u) \text{d} \mathbf u = \int_{\mathcal D_{\mathbf{ \hat u}}} (f(\mathbf u) - \mathcal M(\mathbf u))^2 \,  \text{d} \mathbf u\label{eq:SE}
		\end{align}
		where $f_{ \mathbf{ \hat x}}$ (resp. $\mathcal D_{\mathbf{ \hat x}}$) is the PDF (resp. the support) of the random input vector ${\mathbf{ \hat x}}$ and $f_{\mathbf{ \hat u}}$ (resp. $\mathcal D_{\mathbf{ \hat u}}$) is the PDF (resp. the support) of the random input vector ${\mathbf{ \hat u}}$. The PDF of the i.\,i.\,d. uniform random input vector ${\mathbf{ \hat u}}$ is $f_{\mathbf{ \hat u}}(\mathbf u)=1$. \\
		\\
		The SE value (\autoref{eq:SE}) is generally not known analytically, since the model function $\mathcal M$ is assumed to be known only at certain points as is the case for complex computer models. Therefore, the generalization error is estimated by using a validation set $\{(\mathbf u_{\text{val}, i} ,\, y_{\text{val,}i}),\, \, i = 1\ldots n_\text{val}\}$ with $n_\text{val}$ validation points obtained from evaluations of the computer model. The validation points are obtained by Monte-Carlo sampling with respect to the parameter PDFs. The normalized mean squared error results in
		\begin{equation*}
			\text{NMSE} = \frac{\hat{\text{SE}}}{\hat{\sigma}_y^2} = \frac{1}{\sigma^2_{y_\text{val}}}\frac{1}{n_{\text{val}}}  \sum_{i=1}^{n_{\text{val}}}(y_{\text{val,}i}-\mathcal M(\mathbf u_{\text{val}, i}))^2.
		\end{equation*}
		Here, 
		\begin{align*}
			\overline y &= \frac{1}{n_\text{val}} \sum_{i=1}^{n_\text{val}} y_{\text{val,}i} \quad \text{and}\\
			\sigma^2_{y_\text{val}} &= \frac{1}{n_\text{val}-1}\sum_{i=1}^{n_\text{val}} \left(y_{\text{val,}i}-\overline y\right)^2
		\end{align*}
		are the mean and the variance of evaluations $y_{\text{val,}i}$ as estimates for the output variable $y$, respectively.\\
		\\
		Model accuracy is considered to be high if NMSE values are close to $0$ and low if NMSE values are close to $1$. By definition, values are non-negative and should not exceed $1$ as the covariance between the surrogate model and data would then be higher than the variance of the data. The NMSE error is used since normalization allows comparison between different problems, i.\,e. different scales. Compared to the $Q^2$ error ($Q^2 = 1 - \text{NMSE}$, see e.\,g. \citet{PCEkriging}), the NMSE is more sensible to visualize graphically on a logarithmic scale which is reasonable for small values.\\
		\\
		In the case of expensive computational models, cross-validation techniques such as leave-one-out cross-validation can be applied instead of using a separate validation data set. If the cross-validation error is costly to compute, analytical expressions can be used to reduce computational cost arising by obtaining many separate leave-one-out surrogate models (see e.\,g. \citet{loo_kriging}). Since computationally inexpensive test cases are used in this study to investigate the proposed method, separate validation data sets are used to ensure accurate validation results.

\section{Test cases}\label{sec:test}
	
	The proposed method is applied to several benchmark problems that are shown in \autoref{table:bench}. Probability density functions are assigned to input parameters. All investigated functions contain a non-uniform distribution for at least one input dimension because otherwise the Rosenblatt transformation would not affect the trend function and the proposed method would not differ from the conventional methods.\\

	\begingroup
	\renewcommand{\arraystretch}{1.5}
	\LTcapwidth=\textwidth
	\begin{longtable}{ |c|p{60mm}|c|p{50mm}| } 
		
		\caption{Benchmark functions including labels, mathematical expressions, number of input dimensions, probability density functions (PDF) and parameter correlations.} \label{table:bench}\\
		\hline
		\# & \centering equation & dim. &input parameter PDFs$^\star$ $f_{\hat x_i}$  \newline  and pairwise Pearson\newline  correlation coefficients $\rho_{\hat x_i,\hat x_j}$\newline ($\rho_{\hat x_i,\hat x_j}=0$ if not stated)\\
		\hline
		1 & Oakley \& O'Hagan  \cite{bench_1} \newline $f(x)=5+x+\cos (x)$ & 1 & $\hat x \sim \mathcal N (0,4)$  \\ 
		2 & Lognormal Ratio  \cite{bench_245} \newline $f(\mathbf{x})=\frac{x_{1}}{x_{2}}$& 2 & $\hat x_{1,2} \sim \mathcal {LN} (1,0.5)$ \newline \ $\rho_{\hat x_1,\hat x_2} = 0.3$\\ 
		3 & Webster et al. \cite{bench_3}  \newline$f(\mathbf{x})=x_1^{2}+x_2^{3}$& 2 & $\hat x_1 \sim \mathcal U (1,10)$\newline  $\hat x_2 \sim \mathcal N (2,1)$ \\ 
		4 & Short Column  \cite{bench_245}  \newline$f(\mathbf{x})=1-\frac{4}{1125}\frac{ x_2}{x_1}-\frac{1}{5625}\left(\frac{x_3}{x_1}\right)^2$& 3 & $\hat x_1 \sim \mathcal {LN} (5,0.5) \newline \hat x_2 \sim \mathcal {N} (2000, 400) $\newline$ \hat x_3 \sim \mathcal {N} (500,100)$ \newline $\rho_{\hat x_2,\hat x_3} = 0.5$\\ 
		5 & Cantilever Beam  \cite{bench_245}  \newline$f(\mathbf{x})= \frac{5\cdot 10^5}{x_1} \sqrt{\left(\frac{x_2}{16}\right)^{2}+\left(\frac{x_3}{4}\right)^{2}}$& 3 &  $ \hat x_1 \sim \mathcal {N} (2.9\mathrm{e}7, 1.45\mathrm{e}6)\newline \hat x_2 \sim \mathcal {N} (1000,100) $\newline$ \hat x_3 \sim \mathcal {N} (500,100)$\\ 
		6 & Borehole  \cite{bench_6_1, bench_6_2}  \newline$f(\mathbf{x})=\frac{2 \pi x_3 \left(x_4-x_6\right)}{\ln \left(\frac{x_2}{x_1}\right)\left(1+\frac{2 x_7 x_3}{\ln \left(x_2 / x_1\right) x_1^{2} x_8}+\frac{x_3}{x_5}\right)}$& 8 & $\hat x_1 \sim \mathcal {N} (0.1, 0.0162)\newline \hat x_2 \sim \mathcal {LN} (3700, 4890)
		\newline \hat x_3 \sim \mathcal {U} (63\,070, 115\,600)\newline \hat x_4 \sim \mathcal {U} (990, 1110)$\newline$ \hat x_5 \sim \mathcal {U} (63.1, 116)\newline \hat x_6 \sim \mathcal {U} (700, 820) \newline \hat x_7 \sim \mathcal {U} (1120, 1680)\newline \hat x_8 \sim \mathcal {U} (9\,855, 12\,045)$ \\ 
		7 & Steel Column \cite{bench_7}  \newline  $f(\mathbf{x})=x_1-\frac{P}{2 x_5 x_6}-\frac{x_8 P E_{b}}{x_5 x_6 x_7\left(E_{b}-P\right)}$,\newline
		\text{ }\qquad  $P=x_2+x_3+x_4,\newline
		\text{ }\qquad \,E_{b}=\frac{8\pi^{2}}{9\cdot10^{8}}  x_5 x_6 x_7^{2}x_9 $& 9 & $\hat x_1 \sim \mathcal {LN} (400,35)\newline \hat x_2 \sim \mathcal {N} (5\mathrm{e}5, 5\mathrm{e}4) \newline \hat x_{3,4} \sim \mathcal {G} (6\mathrm{e}5, 9\mathrm{e}4)\newline \hat x_5 \sim \mathcal {LN} (300, 3)$,\newline$ \hat x_6 \sim \mathcal {LN} (20, 2)\newline \hat x_7 \sim \mathcal {LN} (300, 5) \newline \hat x_8 \sim \mathcal {N} (30,10)\newline \hat x_9 \sim \mathcal {W} (2.1\mathrm{e}5, 4200)$ \\ 
		
		8 & Sulfur Model \cite{bench_8}  \newline$f(\mathbf{x})=-5.488\cdot 10^{-9} \newline \text{ }\hspace{1.5cm} \cdot  x_1^{2} x_2 x_3^{2} x_4 x_5 x_6 x_7 x_8 x_9$& 9 & $\hat x_1 \sim \mathcal {LN} (0.76, 0.152)\newline \hat x_2 \sim \mathcal {LN} (0.39, 0.039)\newline \hat x_3 \sim \mathcal {LN} (0.85, 0.085)\newline \hat x_4 \sim \mathcal {LN} (0.3, 0.09)\newline \hat x_5 \sim \mathcal {LN} (5.0, 2.0)$,\newline$ \hat x_6 \sim \mathcal {LN} (1.7, 0.34)\newline \hat x_7 \sim \mathcal {LN} (71.0, 10.65)\newline \hat x_8 \sim \mathcal {LN} (0.5, 0.25)\newline \hat x_9 \sim \mathcal {LN} (5.5, 2.75)$\\ 
		9 & Oakley \& O'Hagan \cite{bench_9} \newline$f(\mathbf{x})=\mathbf{a}_{\mathbf{1}}^{T} \mathbf{x}+\mathbf{a}_{\mathbf{2}}^{T} \sin (\mathbf{x})\newline \text{ } \hspace{1.5cm}+\mathbf{a}_{3}^{T} \cos (\mathbf{x}) +\mathbf{x}^{T} \mathbf{M}\mathbf{x}, \newline \text{ }\hspace{1.5cm} \mathbf a_i, \mathbf M$ according to \cite{bench_9}& 15 &$\hat x_i \sim \mathcal N(0,1), \quad i = 1\ldots 15$\\ 
		\hline
		
		\multicolumn{4}{l}{\thead[l]{${ }^\star$PDF parameters correspond to mean $\mu$ and standard deviation $\sigma$ for normal $\mathcal N$, \\ log-normal $\mathcal {LN}$, Weibull $\mathcal W$ and Gumbel $\mathcal G$ distributions and to lower and upper\\ limit for uniform $\mathcal U$ distributions.}}\\


	\end{longtable}
	\endgroup

	For each problem with number of input dimensions $p$, a number of $n=10 \, p$ training points $\mathbf x_i$ is generated by maximin Latin hypercube sampling. This number is chosen based on the recommendation by \citet{sample_size} for conducting initial experiments. Evaluations of model functions $y_i = f(\mathbf x_i)$ are conducted. Surrogate models are built based on described methods in \autoref{sec:quantile}. The following Gaussian process regression methods are compared: simple kriging, ordinary kriging, universal kriging with linear trend, universal kriging with quadratic trend, universal kriging with transformed linear trend and universal kriging with transformed quadratic trend. A linear trend indicates that linear terms w.\,r.\,t. all input parameters are included. A quadratic trend indicates that polynomial terms up to the order of 2 w.\,r.\,t. all input parameters  are included. A transformed trend indicates that inverse Rosenblatt transformation is applied to the corresponding trend function as demonstrated in \autoref{sec:basis_functions}. Transformations are only applied to linear and quadratic trends because transformation of a zero (simple kriging) or constant (ordinary kriging) trend would not result in any change.\\
	\\
	For all combinations between each surrogate method and each benchmark problem, the demonstrated methods for constructing a surrogate model are applied $10$ times for better statistical validity. As there may be multiple optima or unsuited initial values in the hyperparameter optimization, each optimization run is repeated $20$ times with randomized initial hyperparameters and the hyperparameter set which yields the lowest validation error is selected. Computation time for one experiment therefore includes all $20$ hyperparameter optimization runs. Mean value and standard deviation are determined for the validation errors in all cases.\\
	\\
	The surrogate models are validated according to the validation measures in \autoref{sec:valid}. In this study, a set of $n_\text{val}=1000$ validation points is used for each validation.\\
	\\
	Furthermore, all experiments are conducted for the case where the gradient of the log marginal likelihood is not included and where it is included according to \autoref{eq:log0} and \autoref{eq:log1} for comparison of computation time. For optimization, the L-BFGS-B method is used. In the case without gradient information, the gradient is estimated by 2-point finite difference estimation.\\

\section{Results}\label{sec:results}

\begin{sidewaysfigure}
		\centering
		\includegraphics[width=18cm]{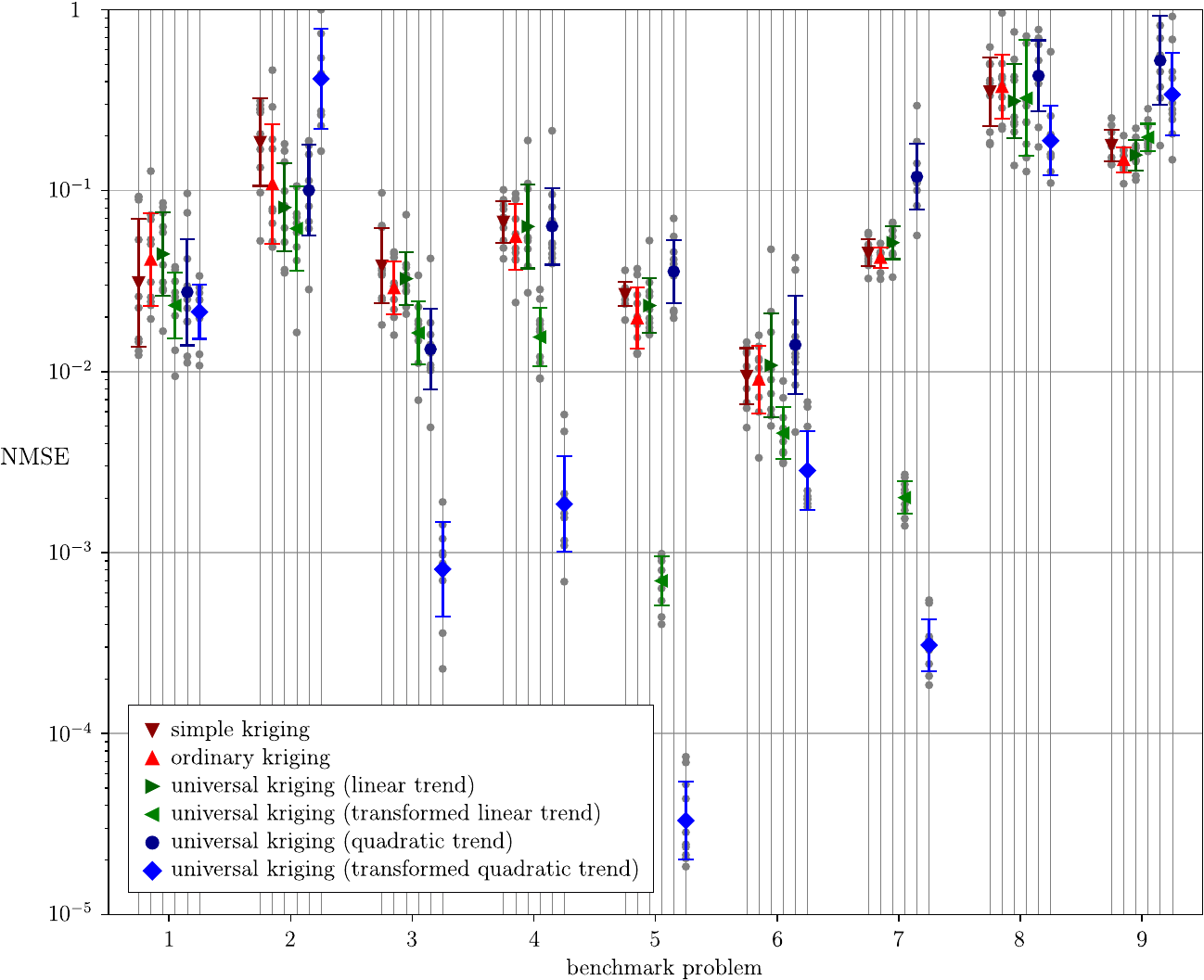}
		\caption{Model validation results: NMSE for all benchmark problems and investigated kriging methods, respectively. Mean values and standard deviations of the 10 experiments are shown for all cases.}
		\label{fig:valid}
	\end{sidewaysfigure}

\begin{figure*}[p!]
		\includegraphics[width=\textwidth]{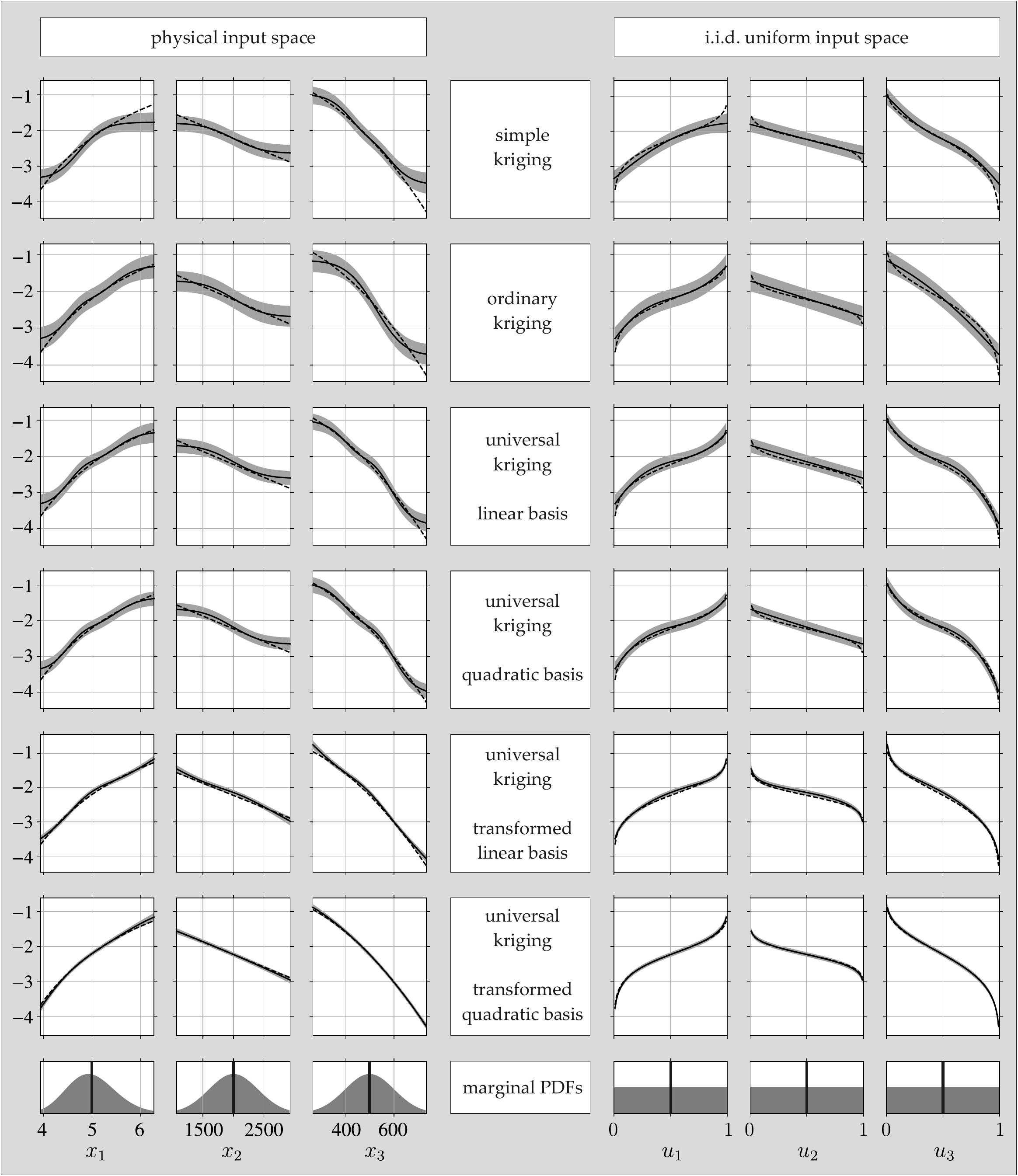}
		\caption{Visualization of benchmark problem $\# 4$ (short column function): True function (dashed line) and kriging prediction mean (solid line) and variance (grey shaded area) for investigated kriging methods (rows) and input parameters (columns), in the  physical (left) and i.\,i.\,d. uniform (right) input space. Only cuts through the hypersurface are shown where one input parameter is changed whereas all other input parameters stay fixed at their mean value (shown in PDF on the bottom).}
		\label{fig:cut}
	\end{figure*}

\begin{sidewaysfigure}
		\centering
		\includegraphics[width=18cm]{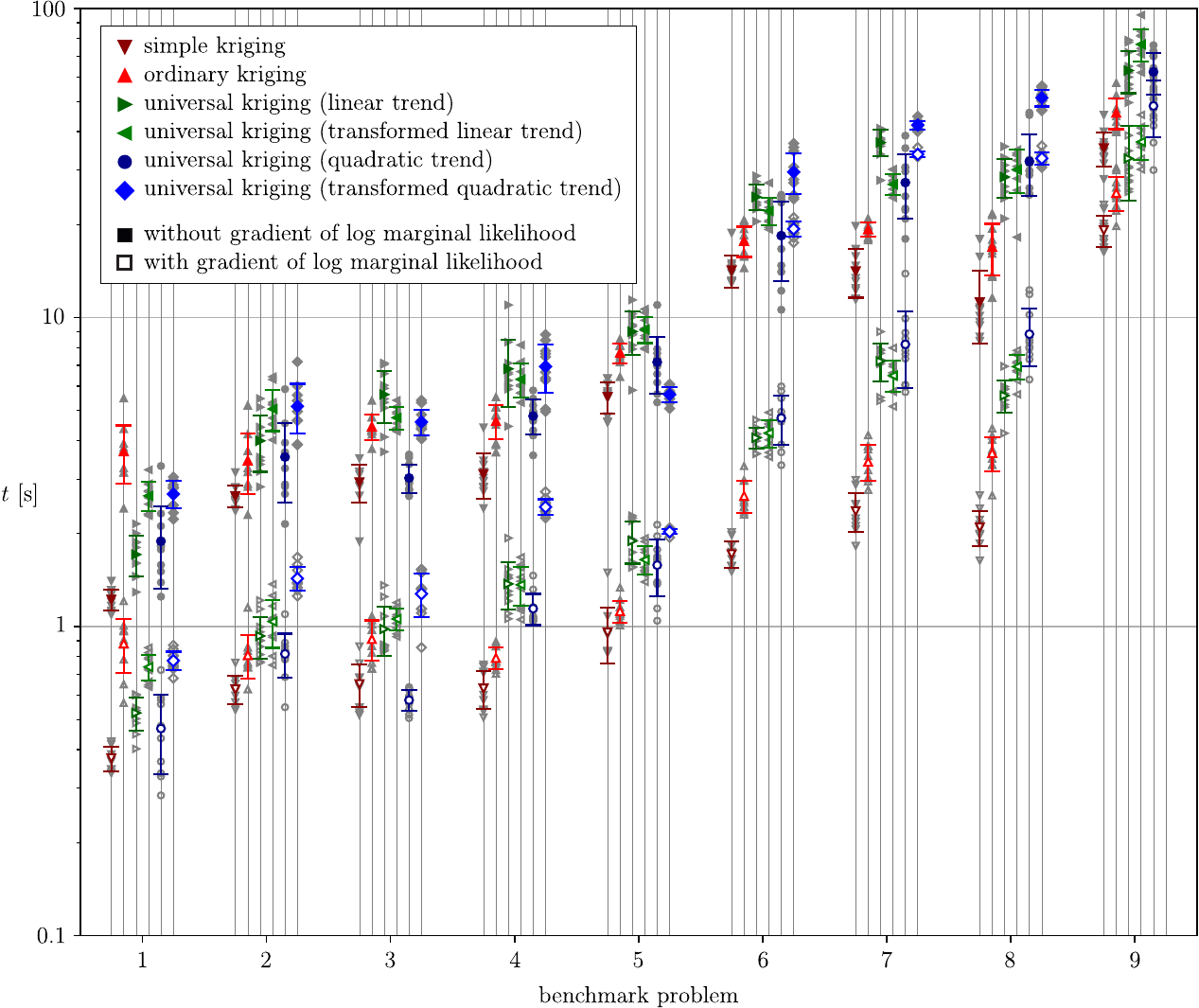}
		\caption{Runtime for obtaining a surrogate model for all benchmark problems and investigated kriging methods, respectively. Results are shown for the cases with (empty markers) and without (filled markers) incorporation of the gradient of the log marginal likelihood for hyperparameter optimization (see \autoref{sec:hyperpar}). Mean values and standard deviations of the 10 experiments are shown for all cases.}
		\label{fig:runtime}
	\end{sidewaysfigure}

	The validation errors for all combinations between benchmark functions and surrogate methods, each consisting of the $10$ experiments, are illustrated in \autoref{fig:valid}, including their mean values and standard deviations. The mean values are shown in \autoref{sec:app_valid}. The simple kriging model shows the highest validation errors on average. The ordinary kriging models yield significantly lower validation errors compared to the simple kriging models. The universal kriging models yield even higher prediction accuracy than simple kriging and ordinary kriging. Depending on the problem, linear or quadratic basis functions are superior with regard to the validation error. However, for most cases a quadratic trend leads to smaller errors. The incorporation of transformed basis functions significantly reduces the validation errors compared to the non-transformed case for both linear and quadratic trends in most cases.\\
	\\
	In \autoref{fig:cut} the effect of incorporating transformed basis functions on the surrogate model is illustrated for the \textit{short column function} (benchmark problem $\#4$). For the purpose of illustration, only cuts through the hypersurface are shown where only one input parameter is changed at a time and the output quantity with respect to each input parameter is shown. The values of the function that is to be predicted as well as the surrogate model are shown in the physical input space and the i.\,i.\,d. uniform input space, respectively.\\
	\\
	In \autoref{fig:runtime} the computation time for all combinations between benchmark functions and surrogate methods are shown including all $10$ experiments with mean values and standard deviations. The computation time is shown for the cases without and with incorporation of the gradient of the log marginal likelihood for hyperparameter estimation. In all cases, the computation time with incorporation of gradient information can be reduced significantly.\\
	\\

\section{Discussion} \label{sec:discussion}

	On average, prediction accuracy increases from simple kriging, to ordinary kriging, to universal kriging, because the surrogate models become more flexible and can better adapt to the problem. The effectiveness of applying certain types of basis functions highly depends on the problem itself. Depending on the relationship between input and output quantities in a model, linear or quadratic basis functions in universal kriging, or both, may lead to an improvement of prediction accuracy. In general, quadratic basis functions offer a more flexible way for the surrogate model to adapt to the data which yields an improvement compared to linear functions. However, polynomials of higher degrees combined with a high number of input dimensions result in a very high number of basis functions, known as the curse of dimensionality, which may lead to  overfitting and high generalization errors.\\
	\\
	The linear and quadratic basis functions that have been transformed by the Rosenblatt transformation lead to a significant improvement compared to linear and quadratic basis functions without transformation in most cases. This is due to the polynomial basis functions being defined in the physical input parameter space rather than in the transformed input parameter space.\\
	\\
	The validation errors are very different between the considered benchmark functions (from RMSE$=0.01$ to $0.4$ for simple kriging) due to different numbers of dimensions and training points and due to incomparable nonlinearities of the problems. For example, model function values of benchmark function $\#2$ become extremely high for small input parameter $x_2$ which is not unlikely according to the PDF. Surrogate models may strongly struggle to capture such relationships as can be seen from large validation errors with wide spread, as shown in \autoref{fig:valid}.\\
	\\
	From \autoref{fig:cut} it becomes clear that linear or quadratic basis functions without proposed transformation yield a linear or quadratic trend in the i.\,i.\,d. uniform input space, respectively. On the other hand, transformed basis functions yield a linear or quadratic trend in the physical input space. As it is generally more natural to assume linear or quadratic trends in the physical parameter space, rather than in the transformed i.\,i.\,d. uniform parameter space, the surrogate models with transformed basis functions are generally more accurate approximations of the model function. This can be further emphasized by the fact that the input parameter transformation generally increases the nonlinearities of the problem. Surrogate models with non-transformed basis functions generally tend to systematically deviate from the model function. In particular, the shape of such surrogate models tends to flatten in regions with lower PDF values (close to the margins) and to steepen in region with high PDF values (close to the mean). This artefact does not occur if transformed basis functions are used that account for the input space transformation.\\
	\\
	The high validation errors for benchmark problem $\#9$ in the case of universal kriging with a quadratic trend indicate the effect of overfitting. Using quadratic multivariate basis functions for a $p$-dimensional problem leads to $p(p+1)/2$ quadratic, $p$ linear and $1$ constant basis functions in a general case. Here, for $p=15$ and given $n=10p$, this means including $136$ basis functions for a problem with $150$ training points. In the context of polynomial chaos expansion, \citet{pol_size} argues that for polynomial fitting the number of training points should be at least $2$-$3$ times as high as the number of basis functions in order to avoid overfitting. If this is not the case, a sparse set of basis functions could be determined, e.\,g. by means of least-angle regression (e.\,g. \citet{PCEkriging}).\\
	\\
	Incorporating the gradient of the log marginal likelihood for hyperparameter optimization leads to a significant computational speedup. The demonstrated equations in \autoref{sec:hyperpar} have therefore been shown to be valid and useful in all cases.

\section{Conclusion} \label{sec:conclude}

	An approach has been proposed where input parameter transformations are taken into account for the construction of basis functions for universal kriging. Since surrogate modeling methods are applied in the i.\,i.\,d. uniform input space to incorporate space-filling designs in a more meaningful way, basis functions in the universal kriging method also have to be defined in this space. It turned out to be more sensible to define basis functions in the physical input space. The inverse Rosenblatt transformation is applied to transform basis functions from the physical input space into the i.\,i.\,d. uniform input space which are then used for universal kriging. The transformed basis functions lead to a significant improvement in prediction accuracy compared to the case of non-transformed basis functions in most cases. Among the benchmark problems, the NMSE is in some cases lower by up to a factor of $10^3$ with transformed basis functions. The authors of this paper highly recommend to consider proposed transformations to basis functions for universal kriging whenever non-uniform input parameter distributions are used and the surrogate model is constructed in the i.\,i.\,d. uniform parameter space. If the input parameters are uncorrelated, the inverse Rosenblatt transformation reduces to quantile functions of the input parameters which simplifies the transformation significantly. To speed up computation, demonstrated equations for gradient-based hyperparameter optimization can be applied.\\
	\\
	The proposed method can be applied to other surrogate modeling methods that incorporate explicit basis functions where the input space is transformed to an i.\,i.\,d. uniform input space, e.\,g. to polynomial regression. In the latter case, the polynomial basis functions are transformed by the inverse Rosenblatt transformation before fitting the regression model to determine the regression coefficients. Eventually, the regression technique is no longer a polynomial regression, because transformations have been applied to the polynomials.\\

\appendix 
\section{Matrix identities}\label{sec:appendix}    
	Derivatives of the elements of an inverse matrix:
	\begin{equation}\label{eq:div1}
		\frac{\partial}{\partial x}\mathbf M(x)^{-1} = -\mathbf M(x)^{-1}\frac{\partial\mathbf M(x)}{\partial x}\mathbf M(x)^{-1}
	\end{equation}
	Derivative of the log determinant of a positive definite symmetric matrix:
	\begin{equation}\label{eq:div2}
		\frac{\partial}{\partial x}\log(|\mathbf M(x)|) = \text{tr}\left(\mathbf M(x)^{-1}\frac{\partial\mathbf M(x)}{\partial x}\right)
	\end{equation}
	Cyclic permutation of matrices in the argument of a trace:
	\begin{equation}\label{eq:spur}
		\text{tr}(\mathbf M_1 \, \mathbf M_2\, \mathbf M_3) = \text{tr}(\mathbf M_2 \,\mathbf M_3\, \mathbf M_1) = \text{tr}(\mathbf M_3\, \mathbf M_1 \,\mathbf M_2)
	\end{equation}

\section{Validation errors}\label{sec:app_valid}
	\renewcommand{\arraystretch}{1.2}
	\begin{table*}[h] 
		\caption{Model validation results: NMSE for all benchmark problems and investigated kriging methods, respectively, as illustrated in \autoref{fig:valid}.}
		\begin{tabular}{ |c|>{\centering\arraybackslash}m{16mm}|>{\centering\arraybackslash}m{16mm}|c| } 
			\hline
			\# & simple kriging & ordinary kriging & \begin{tabular}{c} universal kriging with\\ \hline \begin{tabular}{>{\centering}m{16mm}>{\centering}m{16mm}>{\centering}m{16mm}>{\centering}m{16mm}}linear trend&quadratic trend&transf. linear trend&transf. quadratic trend\end{tabular}\end{tabular} \\  
			\hline
			1	&	4.31E-02	&	4.99E-02	&	\begin{tabular}{>{\centering}m{16mm}>{\centering}m{16mm}>{\centering}m{16mm}>{\centering}m{16mm}}	5.07E-02	&	2.50E-02	&	3.52E-02	&	2.25E-02	\end{tabular}	\\
			2	&	2.10E-01	&	1.49E-01	&   \begin{tabular}{>{\centering}m{16mm}>{\centering}m{16mm}>{\centering}m{16mm}>{\centering}m{16mm}}	9.38E-02	&	6.93E-02	&	1.16E-01	&	5.10E-01	\end{tabular}	\\
			3	&	4.35E-02	&	3.06E-02	&	\begin{tabular}{>{\centering}m{16mm}>{\centering}m{16mm}>{\centering}m{16mm}>{\centering}m{16mm}}	3.47E-02	&	1.76E-02	&	1.53E-02	&	9.41E-04	\end{tabular}	\\
			4	&	6.95E-02	&	6.04E-02	&	\begin{tabular}{>{\centering}m{16mm}>{\centering}m{16mm}>{\centering}m{16mm}>{\centering}m{16mm}}	7.39E-02	&	1.66E-02	&	7.36E-02	&	2.26E-03	\end{tabular}	\\
			5	&	2.71E-02	&	2.13E-02	&	\begin{tabular}{>{\centering}m{16mm}>{\centering}m{16mm}>{\centering}m{16mm}>{\centering}m{16mm}}	2.48E-02	&	7.30E-04	&	3.87E-02	&	3.76E-05	\end{tabular}	\\
			6	&	1.00E-02	&	9.81E-03	&	\begin{tabular}{>{\centering}m{16mm}>{\centering}m{16mm}>{\centering}m{16mm}>{\centering}m{16mm}}	1.40E-02	&	4.85E-03	&	1.72E-02	&	3.27E-03	\end{tabular}	\\
			7	&	4.59E-02	&	4.29E-02	&	\begin{tabular}{>{\centering}m{16mm}>{\centering}m{16mm}>{\centering}m{16mm}>{\centering}m{16mm}}	5.26E-02	&	2.05E-03	&	1.31E-01	&	3.26E-04	\end{tabular}	\\
			8	&	3.83E-01	&	4.11E-01	&	\begin{tabular}{>{\centering}m{16mm}>{\centering}m{16mm}>{\centering}m{16mm}>{\centering}m{16mm}}	3.50E-01	&	4.41E-01	&	4.72E-01	&	2.13E-01	\end{tabular}	\\
			9	&	1.82E-01	&	1.50E-01	&	\begin{tabular}{>{\centering}m{16mm}>{\centering}m{16mm}>{\centering}m{16mm}>{\centering}m{16mm}}	1.60E-01	&	2.00E-01	&	6.18E-01	&	3.93E-01	\end{tabular}	\\
			
			\hline
		\end{tabular}
	\end{table*}

\bibliographystyle{unsrtnat}
\bibliography{references} 

\end{document}